\documentclass[runningheads]{llncs}
\usepackage[T1]{fontenc}
\usepackage{graphicx}
\usepackage{color}

\usepackage{booktabs,makecell, multirow, tabularx}
\usepackage{amssymb}
\usepackage{amsmath}
\usepackage{subcaption}
\usepackage{orcidlink}
\usepackage{bbding}
\begin{document}
\title{DS-HGCN: A Dual-Stream Hypergraph Convolutional Network for Predicting Student Engagement via Social Contagion}
\titlerunning{Dual-Stream Hypergraph Convolutional Network}

\author{Ziyang Fan\inst{1}~\orcidlink{0000-1111-2222-3333} \and
	Li Tao\inst{1(}\Envelope\inst{)}~\orcidlink{0000-0003-0851-1289} \and
	Yi Wang\inst{1}~\orcidlink{0000-0003-3060-9522} \and 
	Jingwei Qu\inst{1}~\orcidlink{0000-0003-4607-8703} \and
	Ying Wang\inst{1}~\orcidlink{0000-0001-7829-7607} \and 
	Fei Jiang\inst{2}~\orcidlink{0009-0001-9495-9903}} 
\authorrunning{Z. Fan et al.}

\institute{Southwest University, Chongqing, 400715, China \\
	\email{tli@swu.edu.cn} 
	\and Chongqing Academy of Science and Technology, Chongqing, China}

%
%


\maketitle  

\begin{abstract}
Student engagement is a critical factor influencing academic success and learning outcomes. Accurately predicting student engagement is essential for optimizing teaching strategies and providing personalized interventions. However, most approaches focus on single-dimensional feature analysis and assessing engagement based on individual student factors. In this work, we propose a dual-stream multi-feature fusion model based on hypergraph convolutional networks (DS-HGCN), incorporating \textbf{social contagion} of student engagement. DS-HGCN enables accurate prediction of student engagement states by modeling multi-dimensional features and their propagation mechanisms between students.The framework constructs a hypergraph structure to encode engagement contagion among students and captures the emotional and behavioral differences and commonalities by multi-frequency signals. Furthermore, we introduce a hypergraph attention mechanism to dynamically weigh the influence of each student, accounting for individual differences in the propagation process. Extensive experiments on a public benchmark dataset demonstrate that our proposed method achieves superior performance and significantly outperforms existing state-of-the-art approaches.
\end{abstract}

\begin{figure}[t]
	\centering
	\includegraphics[width=\linewidth]{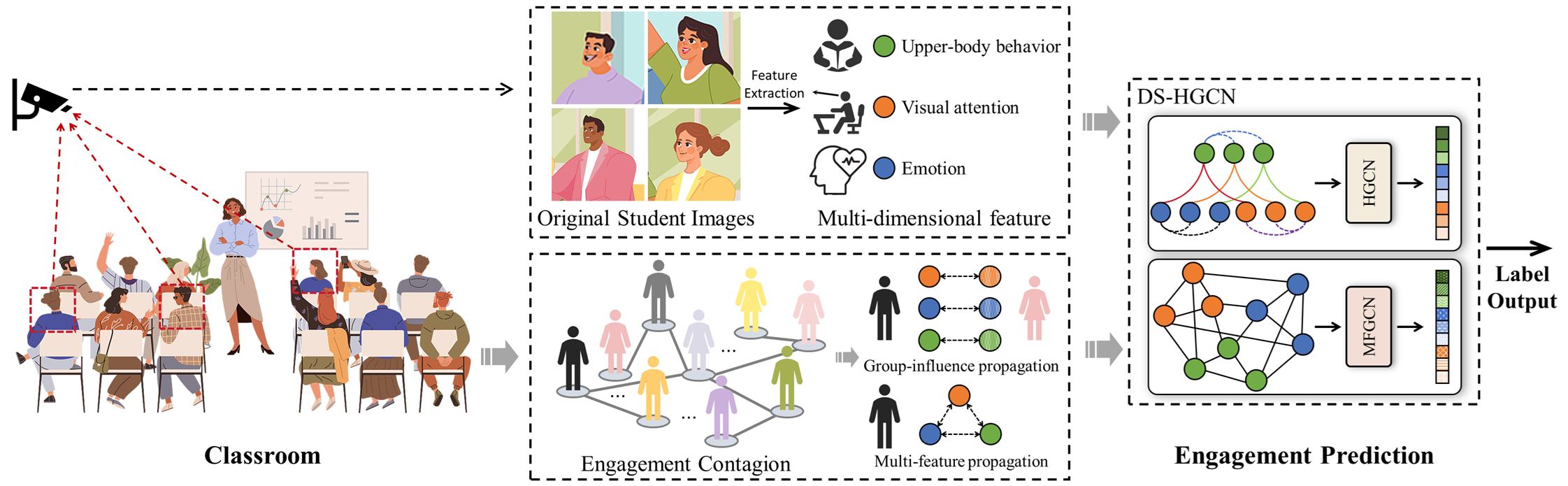}
	\caption{This figure presents the schematic illustration of classroom student engagement prediction. The DS-HGCN leverages the multi-dimensional features and the engagement contagion of students to infer their engagement status.}
	\label{fig:teaser}
\end{figure}

\section{Introduction}
Student engagement is a critical factor influencing academic success and learning outcomes \cite{kuh2001assessing}. It is a multi-dimensional construct encompassing behavioral, emotional, and cognitive aspects, which collectively provide a holistic perspective on a student's participation in the learning process \cite{fredricks2004school}. In modern digital classrooms, accurately predicting student engagement is essential for enhancing teaching strategies and providing personalized interventions.

Given the complexity of engagement, research has progressed from subjective methods like surveys and observations \cite{azevedo2015defining,vallerand1992academic,pekrun2011measuring} to data-driven techniques that use machine learning to analyze diverse data sources, including facial expressions, body language, and physiological signals \cite{azevedo2015defining}. However, these approaches often focus on individual students or unimodal features, providing only a partial understanding. This overlooks the well-documented phenomenon of \textbf{engagement contagion}, where emotions and behaviors spread within a group, significantly influencing individual learning and motivation \cite{mendoza2020social,king2021social}. This leaves a significant research gap, as current models lack the capability to explicitly model the group-level, high-order dependencies inherent in classroom engagement contagion.

To bridge this gap, we introduce the \textbf{Dual-Stream Hypergraph Convolutional Network (DS-HGCN)}, a framework designed specifically to model social contagion for engagement prediction. Our approach leverages a hypergraph to capture the complex, many-to-many relationships within a student group. It operates via two parallel components: a multivariate propagation module to model how engagement spreads among students, and a multi-frequency propagation module to capture the fine-grained distinctions in individual behaviors. To account for varying levels of influence, we further integrate a hypergraph attention mechanism that dynamically assigns weights to these interactions. Our extensive experiments on the RoomReader dataset validate the superiority of our approach. DS-HGCN achieves state-of-the-art accuracy of 94.02\% in binary and 81.37\% in ternary classification, outperforming prior methods by 2.5\% and 5.95\%, respectively. In summary, the main contributions of this work are as follows:

\begin{itemize}
	\item[$\bullet$] We introduce the concept of engagement contagion into a predictive model, using a hypergraph to capture the complex, group-level influences among students in a classroom.
	\item[$\bullet$] We propose a dual-stream architecture, DS-HGCN, that synergistically models both high-order social dependencies and fine-grained feature details.
	\item[$\bullet$] We design a hypergraph attention mechanism that dynamically learns the individual differences in engagement propagation.
\end{itemize}

\section{Related Work}

\subsection{Student Engagement Analysis}
Student engagement analysis is a pivotal research area in educational data mining. Traditional engagement analysis has relied on subjective methods like surveys and direct observation, which lack real-time applicability \cite{azevedo2015defining,vallerand1992academic,pekrun2011measuring}. Consequently, modern approaches leverage machine learning to analyze visible behaviors like attention \cite{kodithuwakku2022emotion,gupta2023facial,khenkar2022engagement,ngoc2019computer}, but their common reliance on a single data modality fails to capture the multifaceted nature of engagement.

To overcome the limitations of unimodal methods, recent research has explored multimodal data fusion. For instance, researchers have combined visual and temporal data (e.g., CNNs and GRUs) to capture emotional dynamics \cite{ding2023online}, integrated various biosignals from wearables to gauge physiological states \cite{sodergaard2023inferring}, and utilized eye-gaze with other signals in VR environments \cite{asish2022detecting}. Others have analyzed engagement through combined behavioral, emotional, and cognitive lenses \cite{vanneste2021computer}. More recently, graph-based models have been proposed to fuse textual, audio, and visual data, offering new perspectives on multimodal analysis \cite{li2024multimodal}. Building on this foundation, our work not only incorporates features related to emotional and behavioral engagement but also models the intricate dependencies between them using a novel graph-based architecture.

\subsection{Graph Neural Networks}
Graph Neural Networks (GNNs) have proven effective for modeling relational data across diverse domains. In educational contexts, they have been applied to classify student behavioral patterns \cite{mubarak2022modeling} and to evaluate engagement by fusing multimodal data sources \cite{li2024multimodal}. 

Standard GNNs are limited to pairwise relationships, whereas classroom interactions are often group-based. Hypergraph Neural Networks (HGNNs) \cite{feng2019hypergraph} are better suited for these high-order relationships, with advancements including dynamic construction \cite{jiang2019dynamic} and attention mechanisms \cite{bai2021hypergraph}. However, a key research gap persists: existing methods overlook \textbf{engagement contagion}. Our work is the first to use a hypergraph to explicitly model this group-level peer influence, addressing a critical limitation in prior research.

\begin{figure*}[t]
	\centering
	\includegraphics[width=\linewidth]{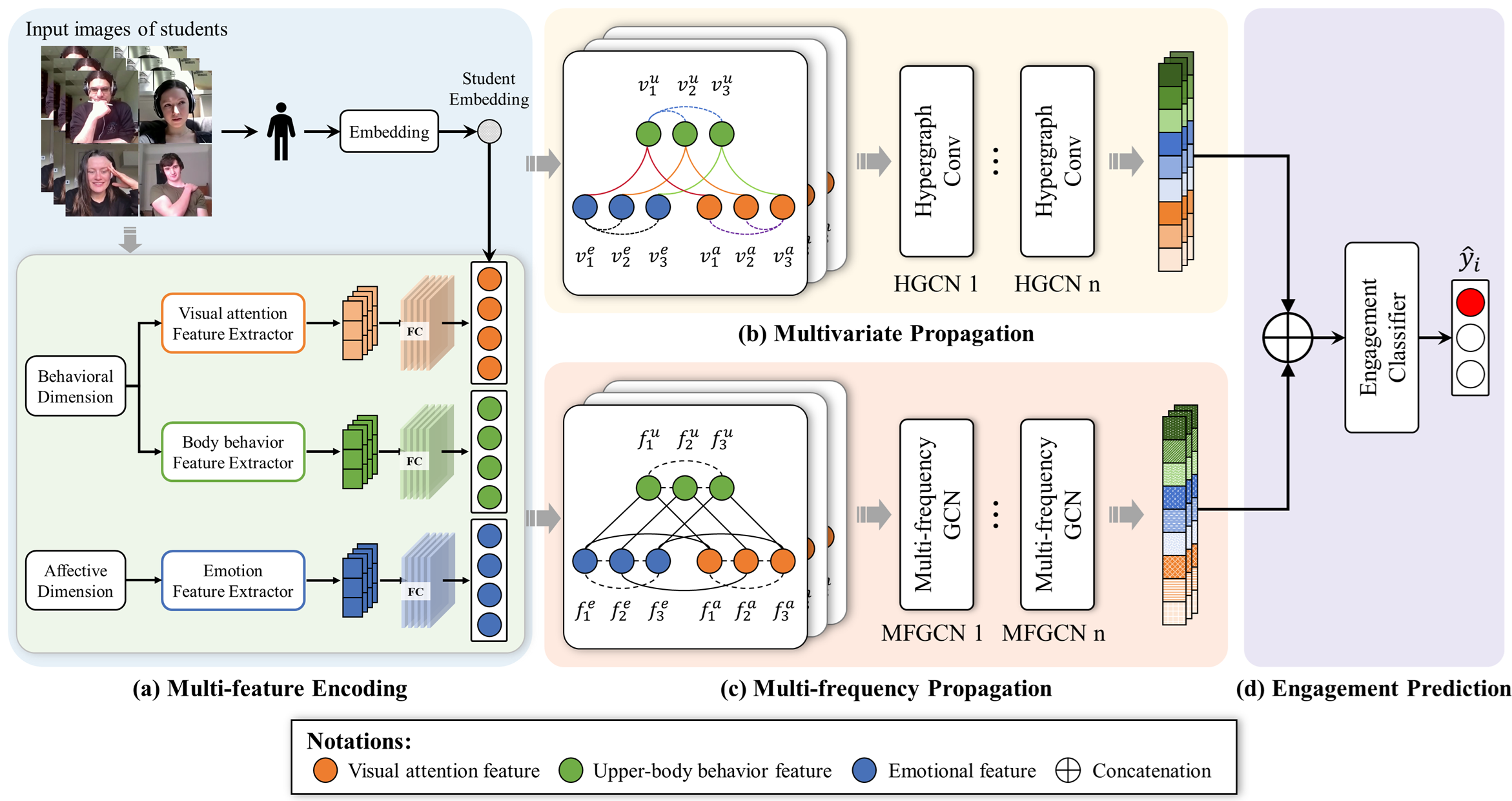}
	\caption{The pipeline of the proposed DS-HGCN.}
	\label{fig:2}
	
\end{figure*}

\section{Method}

The complete pipeline of the proposed DS-HGCN is illustrated in Figure \ref{fig:2}, which consists of four components. To begin with, the Multi-feature Encoder is designed to extract students' multi-dimensional features (Sec. \ref{Feature encoding}). Subsequently, the Multivariate Propagation module models the contagion of engagement among students (Sec. \ref{Multivariate propagation}). In parallel, the Multi-frequency Propagation module captures multi-frequency information from the extracted features (Sec. \ref{Multi-frequency propagation}). At the final stage, the Engagement Classifier predicts engagement levels based on the fused representations (Sec. \ref{Engagement Prediction}).

\subsection{Multi-feature Encoder}{\label{Feature encoding}}

Student engagement consists of behavioral, affective, and cognitive engagement. The multi-feature encoder processes groups of images $\{x_1, x_2, \ldots, x_N\}$, each capturing $N$ students in a classroom at a single moment. For each student $S_i$, the encoder extracts features corresponding to the affective and behavioral dimensions of engagement, which are the primary focus of this vision-based study.

Specifically, we extract three unimodal representations. For emotional features ($x_i^e$), we use a pre-trained DDAMFN++ \cite{zhang2023dual} model, omitting its final classification layer. For visual attentional features ($x_i^a$), we follow prior work \cite{reverdy2022roomreader,li2024multimodal} and use OpenFace \cite{baltruvsaitis2016openface} to obtain gaze direction, head pose, and facial action units. For upper-body behavior features ($x_i^u$), we employ HRNet \cite{lu2024hrnet} to extract human skeleton movements.

These representations are then projected into a unified $D_h$-dimensional space using three separate multilayer perceptrons (MLPs):
\begin{equation} \label{eq:e2}
	c_i^t = W_t x_i^t + b_i^t, \quad t \in \{e, a, u\}
\end{equation}
\noindent where $c_i^t \in \mathbb{R}^{D_h}$. To account for individual differences, we introduce a student-specific embedding $\hat{s}_i$, which is added to each feature representation. The embedding is computed from a one-hot identifier $s_i$:

\begin{equation}\label{eq:e1}
	\hat{s}_i = W_s \ast s_i,
\end{equation}

\noindent where $\hat{s}_i \in \mathbb{R}^{D_h}$ and $W_s$ is a trainable weight matrix. The final encoded representation for each feature is thus:

\begin{equation} \label{eq:e3}
	\hat{c}_i^t = c_i^t + \hat{s}_i,\quad t \in \{e,a,u\}.
\end{equation}

\subsection{Multivariate Propagation}{\label{Multivariate propagation}}
To model engagement contagion, we employ a multivariate propagation module on a hypergraph $\mathcal{H}$. This structure captures heterogeneous feature influences within each student and enables information fusion across different students and dimensions.

\paragraph{\textbf{Hypergraph Construction}}
For a classroom of $N$ students, we construct a hypergraph $\mathcal{H}=(\mathcal{V}_{\mathcal{H}},\mathcal{E}_{\mathcal{H}})$. Its structure is defined by nodes, hyperedges, and an incidence matrix.

\noindent\textbf{Nodes.} The node set $\mathcal{V}_{\mathcal{H}}$ contains $3N$ nodes, where each student $S_i$ is represented by a triplet of nodes $\{v_i^e, v_i^a, v_i^u\}$, corresponding to their emotional, attentional, and upper-body behavior features. The initial representation for each node is the corresponding encoded feature $\{\hat{c}_i^e, \hat{c}_i^a, \hat{c}_i^u\}$ derived from Equation~\ref{eq:e3}.

\noindent\textbf{Hyperedges.} To capture both internal feature interactions and external group influences, we define two types of hyperedges in $\mathcal{E}_{\mathcal{H}}$:

\begin{itemize}
	\item \textbf{Multi-dimensional Hyperedges}: To model the dependencies among a single student's affective and behavioral features, a hyperedge connects the three feature nodes $\{v_i^e, v_i^a, v_i^u\}$ for each student $S_i$. 
	\item \textbf{Group Influence Hyperedges}: To model social contagion, for each feature type $t \in \{e, a, u\}$, a hyperedge connects all corresponding nodes $\{v_1^t, \ldots, v_N^t\}$ across all students.
\end{itemize}
This design, comprising $e\in\mathcal{E}_{\mathcal{H}}(|\mathcal{E}_{\mathcal{H}}|=3+N)$ hyperedges, allows the model to capture high-order correlations beyond simple pairwise interactions.

\begin{figure}[t]
	\centering
	\includegraphics[width=0.7\linewidth]{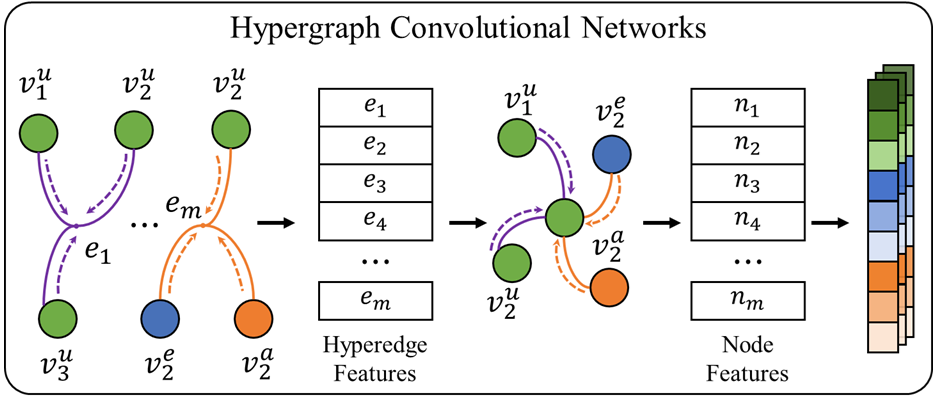}
	\caption{The illustration of the hyperedge convolution layer, which can perform node-edge-node transform, better refining the features using the hypergraph structure.}
	\label{fig:3}
\end{figure}

\noindent\textbf{Incidence Matrix.} The hypergraph's topology is encoded in an incidence matrix $\mathbf{H} \in \mathbb{R}^{|\mathcal{V}_{\mathcal{H}}| \times |\mathcal{E}_{\mathcal{H}}|}$, where each entry indicates if a node belongs to a hyperedge.

\begin{equation}\label{eq:e4}
	H_{ve} = 
	\begin{cases}
		1, & \text{if node } v \in \text{hyperedge } e \\
		0, & \text{otherwise}
	\end{cases}
\end{equation}
This binary matrix serves as a foundation, which will be dynamically weighted by our attention mechanism in \textbf{Hypergraph Attention}.

\paragraph{\textbf{Hypergraph Convolution}}
Information propagation across the hypergraph is achieved via a hypergraph convolution operation, which updates node representations by aggregating information from connected nodes through shared hyperedges. We formulate the hypergraph convolution layer as follows:

\begin{equation}\label{eq:e5}
	\mathbf{Q}^{(l+1)} = \sigma(\mathbf{D}_{\mathcal{H}}^{-1}\mathbf{H}\mathbf{W}_{e}\mathbf{B}^{-1}\mathbf{H}^T\mathbf{Q}^{(l)}\mathbf{P}^{(l)})
\end{equation}

\noindent where $\mathbf{Q}^{(l)} \in \mathbb{R}^{|\mathcal{V}_{\mathcal{H}}|\times D_h}$ is the matrix of node features at layer $l$, and $\sigma$ is a non-linear activation function. $\mathbf{D}_{\mathcal{H}}$ and $\mathbf{B}$ are the diagonal degree matrices for the vertices and hyperedges, respectively. $\mathbf{W}_{e}$ is a diagonal matrix containing learnable weights for each hyperedge, and $\mathbf{P}^{(l)}$ is the trainable weight matrix for the linear transformation at layer $l$.

This operation is fully differentiable, allowing the model to be trained end-to-end via gradient descent. As illustrated in Figure~\ref{fig:3}, the module performs a node-edge-node transformation, effectively refining feature representations based on the high-order relationships encoded in the hypergraph. After $L$ layers of propagation, we obtain the final multivariate representations from the last layer's output $v_{i,(L)}^x$:

\begin{equation}\label{eq:e6}
	\overline{v_{i}^e}=v_{i,(L)}^e, \overline{v_{i}^a}=v_{i,(L)}^a, \overline{v_{i}^u}=v_{i,(L)}^u
\end{equation}

\paragraph{\textbf{Hypergraph Attention}}

To overcome the static nature of the binary incidence matrix $\mathbf{H}$, which limits the model's ability to adapt to varying interaction strengths, we introduce a hypergraph attention mechanism. Inspired by \cite{bai2021hypergraph}, this mechanism dynamically learns the importance of each node $v$ within a hyperedge $e$ by computing an attention score $\gamma_e(v)$.

\begin{equation}\label{eq:e7}
	\gamma_{e_j}(v_i) = \frac{\exp(\sigma(\text{sim}(v_i \mathbf{P}, e_j \mathbf{P})))}{\sum_{k \in \mathcal{N}_v} \exp(\sigma(\text{sim}(v_i \mathbf{P}, e_k \mathbf{P})))}
\end{equation}

\noindent where $\sigma(\cdot)$ is a non-linear activation function and $\mathcal{N}_v$ denotes the neighborhood set of node $v_i$. The similarity function, $\text{sim}(\cdot)$, is defined as a learnable projection:

\begin{equation}\label{eq:e8}
	\text{sim}(v_i,e_j) = a^T[v_i][e_j]
\end{equation}

\noindent where $[\cdot][\cdot]$ denotes concatenation and $\mathbf{a}$ is a trainable weight vector. In this manner, the initial association matrix of the hypergraph is enriched. We represent the attention scores as edge-dependent node weights $\gamma_e(v)$, forming a dynamic weighted association matrix:

\begin{equation}\label{eq:e9}
	H_{ve} = 
	\begin{cases}
		\gamma_e(v), & \text{if node } v \in \text{hyperedge } e \\
		0, & \text{otherwise}
	\end{cases}
\end{equation}

\noindent By making $\mathbf{H}$ in Equation~\ref{eq:e5} dependent on these learnable attention scores, the model can dynamically modulate information flow, effectively capturing the subtle and changing dependencies among student features and group influences during propagation.

\subsection{Multi-frequency Propagation}{\label{Multi-frequency propagation}}
While the multivariate module's aggregation acts as a low-pass filter to capture smoothed dependencies, it risks losing high-frequency details. These details are not noise but often encode critical feature distinctions vital for prediction. To preserve these signals, we introduce a parallel multi-frequency propagation module, inspired by MFGCN \cite{bo2021beyond}, designed to integrate both low- and high-frequency information from student features.

\begin{figure}[t]
	\centering
	\includegraphics[width=0.7\linewidth]{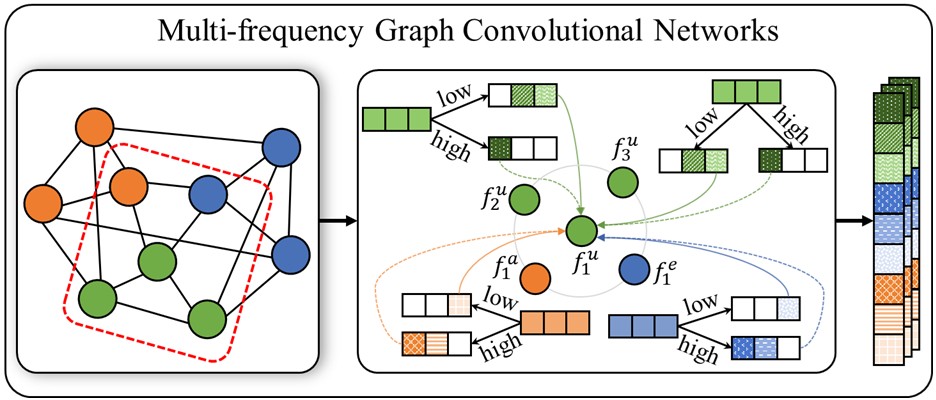}
	\caption{The illustration of the MFGCN is presented. It aggregates low-frequency and high-frequency signals from neighboring nodes.}
	\label{fig:5}
	
\end{figure}

\paragraph{\textbf{Graph Construction}}
We construct an undirected graph $\mathcal{G}=(\mathcal{V}_{\mathcal{G}},\mathcal{E}_{\mathcal{G}})$ that operates in parallel with the hypergraph $\mathcal{H}$. The node set $\mathcal{V}_{\mathcal{G}}$ is identical to $\mathcal{V}_{\mathcal{H}}$, with each node $f_i^t$ ($t \in \{e,a,u\}$) initialized using the encoded representations $\{\hat{c}_i^e,\hat{c}_i^a,\hat{c}_i^u\}$.

Unlike the hypergraph, which uses hyperedges for group dependencies, $\mathcal{G}$ is built on pairwise connectivity to facilitate direct frequency-aware propagation. An edge is created between any two nodes that are connected in the hypergraph structure. This means each node $f_i^t$ is connected to: (1) all nodes $\{f_j^t | j \neq i\}$ representing the same feature from other students, and (2) the other two feature nodes $\{f_i^z|z \neq t\}$ from the same student. The structure is represented by an adjacency matrix $\mathbf{A} \in \mathbb{R}^{|\mathcal{V}_{\mathcal{G}}|\times|\mathcal{V}_{\mathcal{G}}|}$, from which we derive the normalized graph Laplacian:

\begin{equation} \label{eq:e10}
	\mathbf{L}=\mathbf{I}-\mathbf{D}_{\mathcal{G}}^{-1/2}\mathbf{A}\mathbf{D}_{\mathcal{G}}^{-1/2} 
\end{equation} 
\noindent where $\mathbf{D}_{\mathcal{G}}$ is the diagonal degree matrix of $\mathcal{G}$ and $\mathbf{I}$ is the identity matrix. This Laplacian matrix is fundamental for defining our frequency-based filters.

\paragraph{\textbf{Multi-frequency Graph Convolution}}
To explicitly capture signals across the frequency spectrum, we employ two distinct graph filters: a low-pass filter $\mathcal{F}_l$ and a high-pass filter $\mathcal{F}_h$. Following the design in \cite{bo2021beyond}, these are defined based on the graph structure:
\begin{equation}\label{eq:e11}
	\begin{aligned} 
		\mathcal{F}_l &= \mathbf{I} + \mathbf{D}_{\mathcal{G}}^{-1/2}\mathbf{A} \mathbf{D}_{\mathcal{G}}^{-1/2} = 2\mathbf{I} - \mathbf{L} \\
		\mathcal{F}_h &= \mathbf{I} - \mathbf{D}_{\mathcal{G}}^{-1/2}\mathbf{A} \mathbf{D}_{\mathcal{G}}^{-1/2} = \mathbf{L}
	\end{aligned}
\end{equation}
The low-pass filter $\mathcal{F}_l$ averages features from neighbors, promoting smoothness, while the high-pass filter $\mathcal{F}_h$, being the graph Laplacian itself, amplifies the differences between a node and its neighbors, capturing high-frequency distinctions.

We then integrate these filters into a unified propagation layer where the model learns to adaptively combine the low- and high-frequency components. The feature update rule for layer $k$, which takes the feature matrix $\mathbf{F}^{(k)}$ as input, is:
\begin{equation}\label{eq:e12}
	\mathbf{F}^{(k+1)} = \sigma \left( (\mathcal{F}_l \mathbf{F}^{(k)}) \mathbf{W}_{l}^{(k)} + (\mathcal{F}_h \mathbf{F}^{(k)}) \mathbf{W}_{h}^{(k)} \right)
\end{equation}
\noindent Here, $\mathbf{F}^{(k)}$ is the matrix of node features $f_{i,(k)}^t$ at layer $k$. $\mathbf{W}_{l}^{(k)}$ and $\mathbf{W}_{h}^{(k)}$ are learnable weight matrices that adaptively balance the influence of low- and high-frequency signals, and $\sigma$ is a non-linear activation function.

By stacking $K$ such layers, the module iteratively propagates and refines the multi-frequency information. After $K$ layers of message passing, the final frequency-aware engagement representations are obtained as:
\begin{equation}\label{eq:e13}
	\overline{f_{i}^e}=f_{i,(K)}^e, \overline{f_{i}^a}=f_{i,(K)}^a, \overline{f_{i}^u}=f_{i,(K)}^u
\end{equation}

\subsection{Engagement Prediction}{\label{Engagement Prediction}}
After obtaining representations from both the multivariate and multi-frequency modules, the final step is to predict the engagement state for each student, categorized into one of $C$ predefined classes. To achieve this, we first form a comprehensive representation $\mu_i$ for each student $S_i$ by concatenating the outputs from both propagation streams:

\begin{equation}\label{eq:e17}
	\mu_i = \overline{v_{i}^e} \oplus \overline{f_{i}^e} \oplus \overline{v_{i}^a} \oplus \overline{f_{i}^a} \oplus \overline{v_{i}^u} \oplus 
	\overline{f_{i}^u} 
\end{equation}

\noindent where $\mu_i$ serves as a comprehensive feature representation, capturing engagement contagion and multi-frequency interactions relevant to engagement prediction.

The engagement classification is then performed by passing $\mu_i$ through a transformation layer followed by a softmax function:

\begin{equation}\label{eq:e18} 
	\begin{aligned}
		\mathcal{P}_i &= \text{softmax}(W_4 \cdot \text{ReLU}(\mu_i) + b_4), \\
		\hat{y}_i &= \arg\max_{\tau}(\mathcal{P}_i[\tau]),
	\end{aligned}
\end{equation}

\noindent where $W_4$ is a trainable weight matrix, $\mathcal{P}_i \in \mathbb{R}^C$ represents the probability distribution over engagement categories, and $\hat{y}_i$ is the predicted engagement label for student $S_i$.

The model is trained end-to-end by minimizing the standard categorical cross-entropy loss, averaged over all student samples in the training set $\mathcal{D}$, with L2 regularization to prevent overfitting. The loss function is defined as:
\begin{equation}\label{eq:e19}
	\mathcal{L} = - \frac{1}{|\mathcal{D}|} \sum_{i \in \mathcal{D}} \log \mathcal{P}_{i}[y_{i}] + \lambda \|\theta\|_2^2
\end{equation}
\noindent where $y_i$ is the ground-truth label for student $S_i$, $\mathcal{P}_{i}[y_{i}]$ is the model's predicted probability for that true class, $\theta$ represents all trainable parameters of the model, and $\lambda$ is the hyperparameter controlling the strength of the L2 regularization.

\section{Experimental Results}{\label{Result}}
In this section, we conduct a series of comprehensive experiments to rigorously evaluate the effectiveness and robustness of our proposed DS-HGCN. We first detail the experimental setup in Section~\ref{Experimental Setup}. We then demonstrate the superiority of our model by comparing it against state-of-the-art methods in Section~\ref{Baseline Comparisons}. To validate our design choices, we present in-depth ablation studies in Section~\ref{Ablations} to analyze the contribution of each core component. Finally, we provide further analysis on hyperparameter sensitivity and data efficiency in Section~\ref{sec:further_analysis} to offer a holistic assessment of the model.

\subsection{Experimental Setup}{\label{Experimental Setup}}

\paragraph{\textbf{Data Preparation.}}
We evaluate our method on the RoomReader dataset~\cite{reverdy2022roomreader}, a benchmark for student engagement analysis. Following data filtering and frame extraction, we address the dataset's significant class imbalance as noted in prior work~\cite{li2024multimodal}. Following established protocols~\cite{li2024multimodal,sumer2021multimodal}, we formulate two tasks from the continuous annotations: a \textbf{binary} task ('Engaged' vs. 'Disengaged') and a \textbf{ternary} task ('low' vs. 'medium' vs. 'high' engagement), resulting in a final evaluation set of 1,252 unique classroom snapshots.

\paragraph{\textbf{Implementation Details}} Our model is implemented in PyTorch and trained on a single NVIDIA 3090 GPU. We use the Adam optimizer with an initial learning rate of $1 \times 10^{-5}$. A dropout rate of 0.5 is applied to the network to mitigate overfitting. To address the remaining class imbalance, we incorporate class weights into the cross-entropy loss function during training. The number of hypergraph convolution layers ($L$) and graph convolution layers ($K$) were set to 3 and 3, respectively. Model performance is evaluated using Accuracy, F$_1$-score, and Area Under the Curve (AUC).

\subsection{Baseline Comparisons}{\label{Baseline Comparisons}}
To validate the effectiveness of our proposed DS-HGCN, we conducted comprehensive comparisons against several representative and state-of-the-art engagement prediction models. For a fair evaluation, We reimplemented these baselines using the RoomReader dataset, adhering to the specified data formats. The baselines include: 

\begin{itemize}
\item[$\bullet$] \textbf{Single-Feature Models:} This category benchmarks conventional, non-graph-based approaches that model engagement from a single feature. Key examples include: \textbf{TCCT-NET}~\cite{vedernikov2024tcct}, which analyzes behavioral signals via a two-stream fusion network; \textbf{EnsModel}~\cite{thong2019engagement}, which uses a deep network to predict engagement from facial behaviors; \textbf{ConvLSTM}~\cite{del2020you}, which processes video frames with an integrated convolutional regression model; \textbf{ED-MTT}~\cite{copur2022engagement}, which improves accuracy through a multi-task training framework; and \textbf{EG-NET}~\cite{mohamad2019automatic}, which applies a standard CNN for facial expression analysis.

\item[$\bullet$] \textbf{Multi-Feature Model:} To demonstrate the benefit of incorporating multiple features, we compared \textbf{Bootstrap}~\cite{wang2019bootstrap}, a method that combines both head and body features, representing a step beyond single-feature analysis.

\item[$\bullet$] \textbf{Graph-based Model (SOTA):} Finally, to position our work against the direct state-of-the-art, we benchmark against \textbf{Haar-MGL}~\cite{li2024multimodal}. This model employs a multimodal graph learning framework to fuse text, audio, and image data, providing the most relevant and powerful point of comparison for graph-based engagement prediction.
\end{itemize}

\begin{table}[t]
	\caption{Comparison with previous state-of-the-art methods on the RoomReader dataset. Entries marked with $^\diamond$ are taken from paper \cite{li2024multimodal}, whereas those with $^\dagger$ come from our re-implementation using open-source codes.}
	
	\centering
	\resizebox{0.95\textwidth}{!}{
		\begin{tabular}{l|ccc}
			\midrule
			\multirow{2}{*}{\textbf{Methods}} & Binary Classification & Ternary Classification 
			& \multirow{2}{*}{\begin{tabular}[c]{@{}c@{}}Feature Dimension\\Type \end{tabular}} \\
			& Acc.(\%) & Acc.(\%) &  \\ \midrule
			TCCT-NET$^\dagger$ \cite{vedernikov2024tcct} & 73.99 ± 1.43 & 60.48 ± 1.18 & Single \\
			EnsModel$^\dagger$ \cite{thong2019engagement} & 75.30 ± 3.50 & 69.53 ± 3.54 & Single \\
			ConvLSTM$^\dagger$ \cite{del2020you} & 76.50 ± 1.85 & 74.53 ± 1.50 & Single \\
			EG-NET$^\dagger$ \cite{mohamad2019automatic} & 77.38 ± 1.27 & 72.76 ± 2.24 & Single \\
			ED-MTT$^\dagger$ \cite{copur2022engagement} & 73.80 ± 3.35 & 71.20 ± 1.35 & Single \\
			Bootstrap$^\dagger$ \cite{wang2019bootstrap} & 82.42 ± 2.15 & 75.42 ± 2.43 & Multiple \\
			Haar-MGL$^\diamond$ \cite{li2024multimodal} & 90.18 ± 1.34 & - & Multiple \\ \midrule
			\textbf{DS-HGCN(Ours)} & \textbf{94.02 ± 0.49} & \textbf{81.37 ± 0.34} & Multiple \\ \midrule
	\end{tabular}}
	\label{tab:baseline}
\end{table}

Table~\ref{tab:baseline} presents the quantitative results, averaged over five independent trials with an 80\%/20\% training/testing split. As shown, DS-HGCN significantly outperforms all baselines across both binary and ternary classification tasks. The comparison reveals two key insights. First, multi-feature methods consistently surpass single-feature models, confirming the value of a holistic behavioral analysis. Second, DS-HGCN establishes a new state-of-the-art, outperforming the strong graph-based Haar-MGL. This superiority highlights the unique advantage of our dual-stream hypergraph network in explicitly modeling higher-order social contagion—a critical dynamic overlooked by previous approaches.

\begin{table*}[t]
	\caption{Ablation studies of DS-HGCN. We compare our method with a collection of variants described in Sec.\ref{Ablations}}
	\centering
	\resizebox{0.95\textwidth}{!}{
	\begin{tabular}{ll|>{\centering\arraybackslash}p{1.3cm}>{\centering\arraybackslash}p{1.2cm}>{\centering\arraybackslash}p{1.2cm}|>{\centering\arraybackslash}p{1.3cm}>{\centering\arraybackslash}p{1.2cm}>{\centering\arraybackslash}p{1.2cm}}
		\midrule
		& \multirow{2}{*}{\textbf{Candidates}} & \multicolumn{3}{c|}{Binary Classification Task} & \multicolumn{3}{c}{Ternary Classification Task} \\
		& \multicolumn{1}{c|}{} & Acc. (\%) &  $\text{F}_1$ & AUC & Acc. (\%) & $\text{F}_1$ & AUC \\ \midrule
		\textbf{} & \textbf{DS-HGCN(full)} & \textbf{94.02} & \textbf{0.939} & \textbf{0.893} & \textbf{81.37} & \textbf{0.816} & \textbf{0.921} \\ \midrule
		1 & w/o multivariate info. & 89.84 & 0.891 & 0.845 & 75.70 & 0.755 & 0.885 \\
		2 & w/o multi-frequency info. & 92.13 & 0.917 & 0.857 & 77.29 & 0.776 & 0.904 \\ \midrule
		3 & w/o visual attention features. & 84.50 & 0.832 & 0.805 & 68.45 & 0.685 & 0.843 \\
		4 & w/o body behavior features. & 86.37 & 0.859 & 0.823 & 73.37 & 0.742 & 0.862 \\
		5 & w/o emotion features. & 81.56 & 0.803 & 0.784 & 62.77 & 0.631 & 0.784 \\ \midrule
		6 & w/o Hypergraph Attention. & 92.73 & 0.925 & 0.872 & 79.23 & 0.793 & 0.867 \\ \midrule
	\end{tabular}}
	\label{tab:Ablations}
\end{table*}

\subsection{Ablation Study}{\label{Ablations}}

Our ablation study (Table~\ref{tab:Ablations}) validates the contribution of each component in DS-HGCN. Removing the multivariate propagation stream (Variant 1) causes the most significant accuracy drop (4.18\%/5.67\% in binary/ternary tasks), confirming that modeling social contagion via hypergraphs is the primary driver of our model's success. The multi-frequency stream (Variant 2) is also crucial, as its absence markedly degrades performance. An analysis of input features (Variants 3-5) reveals that emotional cues are indispensable, with their removal causing the sharpest performance decline (12.46\% in binary accuracy). Finally, disabling hypergraph attention (Variant 6) impairs performance, validating our dynamic weighting scheme. These results affirm that each design choice is integral to our model's state-of-the-art performance.

\subsection{Further Analysis}\label{sec:further_analysis}
\begin{figure}[t]
	\centering
	\begin{subfigure}[b]{0.45\textwidth}
		\includegraphics[width=\textwidth]{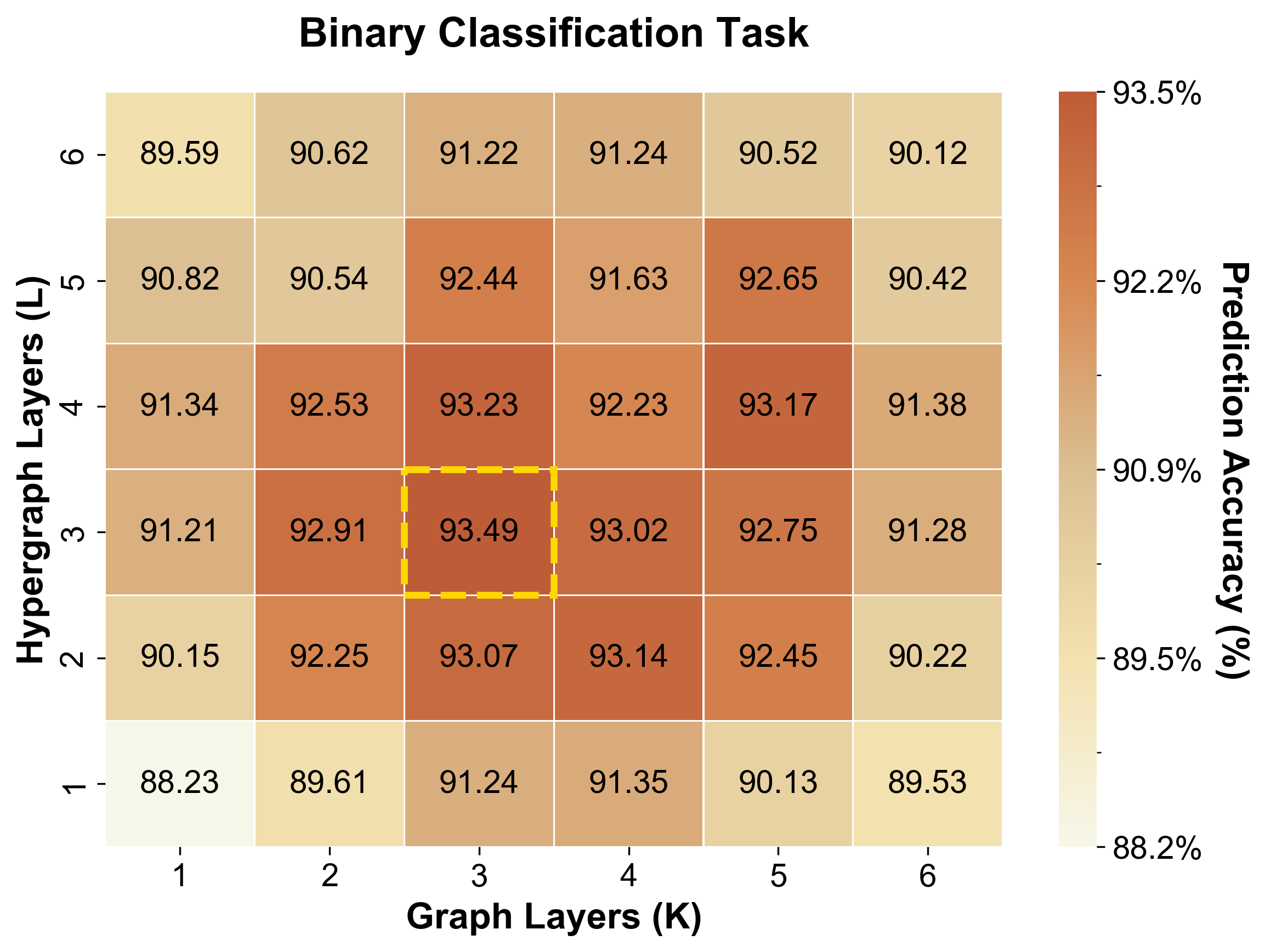} 
		\label{fig:hyperparams1}
	\end{subfigure}
	\begin{subfigure}[b]{0.45\textwidth}
		\includegraphics[width=\textwidth]{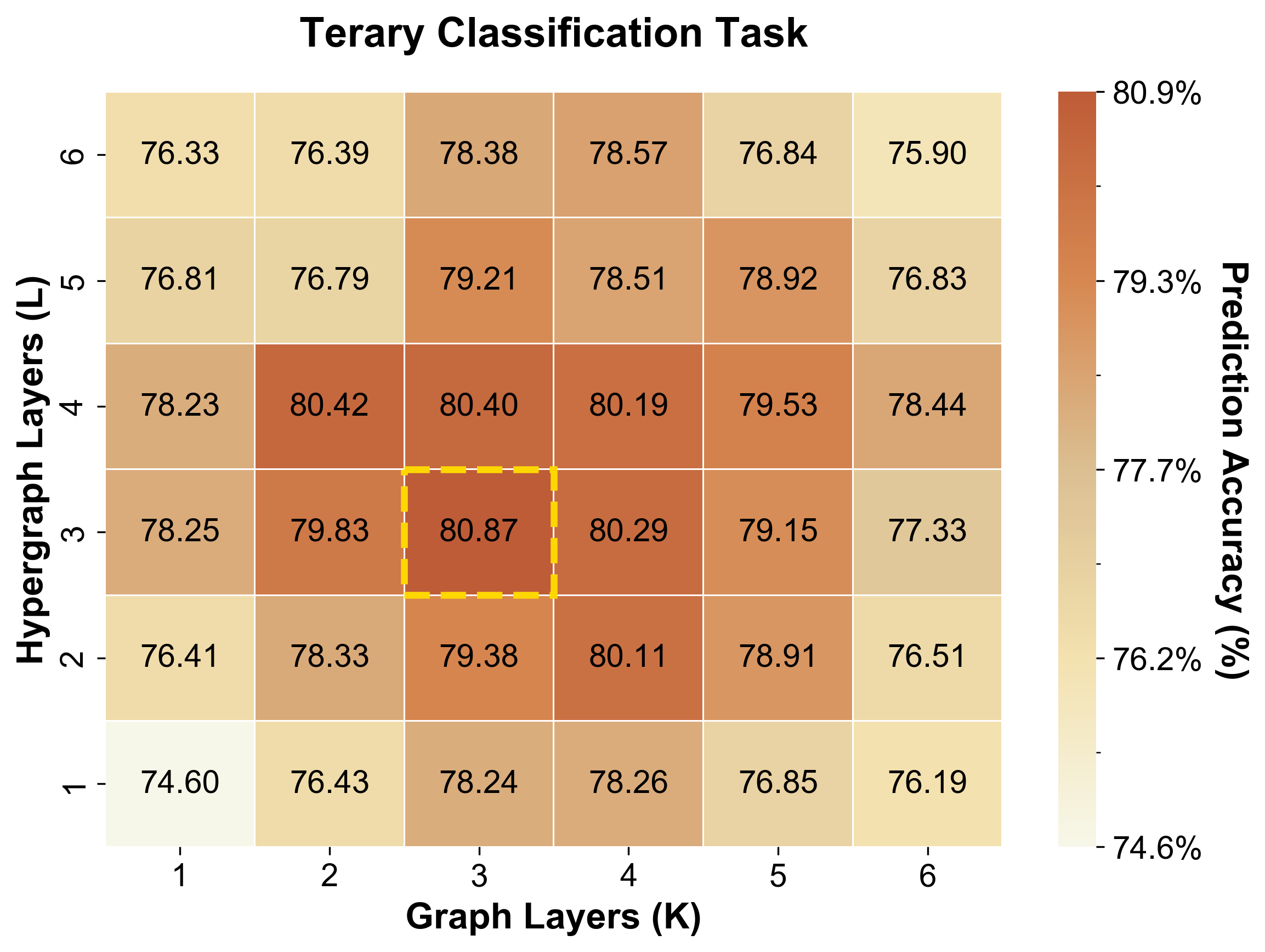} 
		\label{fig:hyperparams2}
	\end{subfigure}
	\caption{Heatmaps illustrating the performance on binary and ternary classification tasks with varying numbers of hypergraph layers ($L$) and graph layers ($K$).}
	\label{fig:hyperparams}
\end{figure}

\begin{figure}[t]
	\centering
	\begin{subfigure}[b]{0.45\textwidth}
		\includegraphics[width=\textwidth]{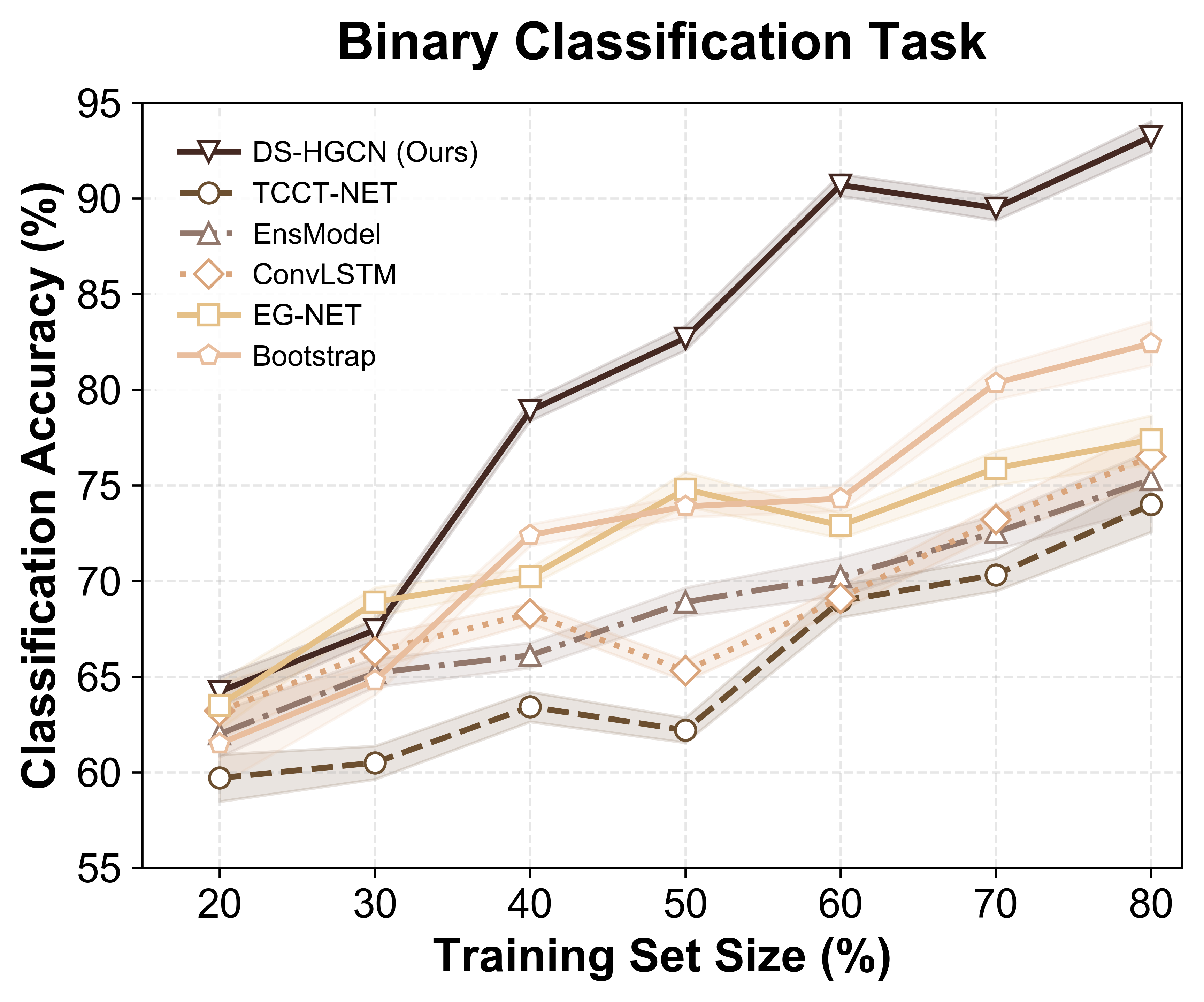} 
		\label{fig:data_scale1}
	\end{subfigure}
	\begin{subfigure}[b]{0.45\textwidth}
		\includegraphics[width=\textwidth]{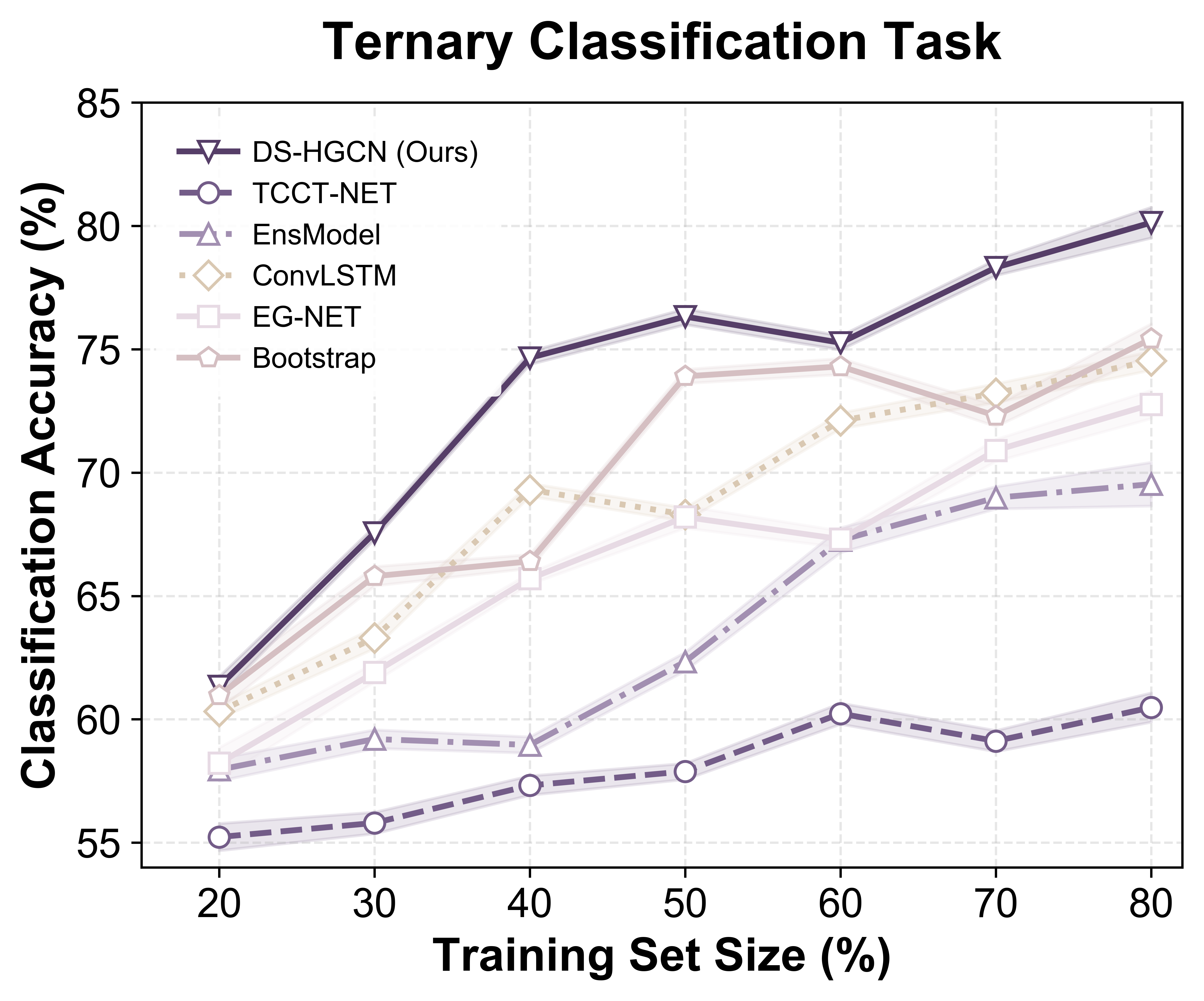} 
		\label{fig:data_scale2}
	\end{subfigure}
	\caption{Comparing models' performance under varying training data proportions for both classification tasks. Shaded area indicates standard deviation.}
	\label{fig:data_scale}
\end{figure}

\paragraph{\textbf{Discussions on graph layers.}} We analyzed the impact of model depth by varying the number of hypergraph ($L$) and graph ($K$) layers from 1 to 6. As illustrated in Figure~\ref{fig:hyperparams}, performance peaks at $L=3$ and $K=3$ for both classification tasks. This suggests a depth of 3 is optimal for capturing engagement contagion without causing over-smoothing, as deeper models yield negligible gains.
\paragraph{\textbf{Impact of Training Data Scale.}} To assess data efficiency, we trained our model on subsets of the training data, ranging from 20\% to 80\%. Figure~\ref{fig:data_scale} shows that DS-HGCN consistently outperforms baselines, even in low-data regimes. Notably, its smaller and more stable standard deviation across all data scales confirms its superior robustness and generalization capability.
\paragraph{\textbf{Analysis of Computational Complexity.}} We evaluated the overhead of our hypergraph attention (HA) mechanism. Table~\ref{tab:complexity} shows that the full DS-HGCN model introduces only a marginal increase in parameters (3.076M to 3.079M) and a slight rise in training time compared to a version without attention. Crucially, the inference time remains highly efficient (8.52 ms/batch), making our approach practical for applications like generating periodic engagement reports.

\begin{table}[h]
	\caption{Comparison of computational costs and efficiency.}
	\centering
	\resizebox{0.78\textwidth}{!}{
		\begin{tabular}{l|c|c|c}
			\toprule
			\textbf{Model}  & \textbf{\makecell{Parameters \\ (M)}} & \textbf{\makecell{Training Time \\ (s/epoch)}} & \textbf{\makecell{Inference Time \\ (ms/batch)}} \\
			\midrule
			DS-HGCN (w/o HA) & 3.076 & 3.69  & 7.31 \\
			\textbf{DS-HGCN (full)} & 3.079 & 4.39  & 8.52 \\
			\bottomrule
	\end{tabular}}
	
	\label{tab:complexity}
\end{table}

\section{Conclusion}
In this paper, we introduced DS-HGCN, a dual-stream hypergraph convolutional network that models social contagion to predict student engagement. Our approach employs parallel hypergraph and graph convolutional networks to explore high-order and complex relationships among various student features and fully leverage the correlations and dependencies of features across different frequency information. Extensive experiments on the RoomReader dataset demonstrate the superiority of our method, achieving 94.02\% accuracy in binary and 81.37\% in ternary classification. While our image-based approach is effective, future work could enhance it by incorporating other modalities like audio and text or exploring proxies for cognitive engagement. Ultimately, DS-HGCN highlights the importance of social context in learning analytics and provides a powerful tool for automated engagement monitoring.

\newpage

\bibliographystyle{splncs04}
\bibliography{conf}
\end{document}